\documentclass[aps,prc,twocolumn,showpacs,groupedaddress,floatfix]{revtex4}

\usepackage{graphicx}
\usepackage{dcolumn}
\usepackage{amsfonts}

\begin{document}

\title{3-D unrestricted TDHF fusion calculations using the full Skyrme interaction}

\author{A.S. Umar and V.E. Oberacker}
  \affiliation{Department of Physics and Astronomy, Vanderbilt University, 
      Nashville, Tennessee 37235, USA}

\date{\today}


\begin{abstract}
We present a study of fusion cross sections using a new generation Time-Dependent 
Hartree-Fock (TDHF) code which contains no approximations regarding collision geometry
and uses the full Skyrme interaction, including all of the time-odd terms. In addition, the code uses
the Basis-Spline collocation method for improved numerical accuracy. A comparative
study of fusion cross sections for  $^{16}O + ^{16,28}O$  is made with the older TDHF results
and experiments. We present results using the modern Skyrme forces
and discuss the influence of the new terms present in the interaction.
\end{abstract}
\maketitle


\section{\label{sec:intro}Introduction}

With the increasing availability of radioactive ion-beams~\cite{DOE02} the study of
structure and reactions of exotic nuclei are now possible~\cite{Li03,Li05,Ji04}.
The microscopic description
of such nuclei will lead to a better understanding of the interplay among the strong, Coulomb,
and the weak interactions as well as the enhanced correlations present in these many-body
systems. This has lead to a considerable theoretical effort to perform nuclear structure
calculations with ever increasing accuracy and extensive investigations of the nuclear effective
interaction~\cite{BHR03}.

From a theoretical point of view, these highly complex many-body systems are often described in
macroscopic terms. This has been particularly true in the case of non-relativistic heavy-ion
collisions~\cite{Ni77}. For example, the time evolution of the nuclear surface and the
corresponding geometrical shape provides a very useful parameter to help organize experimental
data. Using this approach numerous evolutionary models have been developed to explain
particular aspects of the experimental data~\cite{Hu85,Ra79,FG04}. These methods provide a useful
and productive means for quantifying multitudinous reaction data. In practice, they require a
quantitative understanding of the data as well as a clear physical picture of the important
aspects of the reaction dynamics. The depiction of the collision must be given at the onset,
including the choice of coordinates which govern the evolution of the reaction.  Guessing the
correct degrees of freedom is extremely hard, without a full understanding of the dynamics,
and can easily lead to misbegotten results.  More importantly, it is often not possible
to connect these  macroscopic classical parameters, describing nuclear matter under extreme
excitation and rearrangement, with the more fundamental properties of the nuclear force.
Ultimately, these difficulties can only be overcome with a fully microscopic theory of the
collision dynamics.

In this paper, we utilize the time-dependent Hartree-Fock (TDHF) method.
It is generally acknowledged that the TDHF method provides a
useful foundation for a fully microscopic many-body theory of low-energy heavy-ion reactions
\cite{Ne82,DDKS,Sv79}. The TDHF method is most widely known in nuclear physics in
the small amplitude domain, where it provides a useful description of collective states
\cite{Be75,US86, UmOb05}, and is based on the mean-field formalism which has been a relatively
successful approximation to the nuclear many-body problem for reproducing the principal
properties of stable nuclei throughout the periodic table. During 1970's and 1980's the TDHF
theory has been widely used in the study of fusion excitation functions, fission,
deep-inelastic scattering of heavy mass systems, and nuclear molecular resonances
\cite{Ne82,DDKS,US85}, while providing a natural foundation for many other studies. An
account of some of the previous TDHF applications can be found in Refs.~\cite{Ne82,DDKS}.

In the next section we will summarize some theoretical aspects of TDHF theory and give
an account of earlier calculations as it is relevant to this work. In Section~\ref{sec:fuscross}
we present new TDHF fusion calculations and compare them to older results and, when available,
experiments.


\section{\label{sec:formal}Theoretical details}

Despite its wide usage, it has been difficult to assess the reliability of the TDHF
calculations due to an occasional imperfect or even incorrect reproduction of experimental
behavior. This has naturally lead to consider various extensions to the theory, particularly
the inclusion of the two-body collisions~\cite{WoDa,Toh,TU02a}. However, there are important components of the basic
theory which have not yet been fully implemented, and the viability of the analysis depends on
the overall accuracy of the TDHF calculations. The assumptions and approximations that may
impact the results of the TDHF calculations can be categorized as:
(a) Symmetry assumptions about the collision dynamics,
(b) symmetry assumptions used for the nuclear force,
(c) accuracy of the numerical implementation.
Approximations of any type limit the number of degrees of freedom accessible during a collision,
and hence the nature and degree of dissipation~\cite{USR,KrLa,RU,US89}. The understanding of 
the dissipative mechanisms in the TDHF theory is vital for establishing the region of validity of the
mean-field approximation and providing estimates for the importance of the mean-field effects at
higher energies. In TDHF, the dissipation of the translational kinetic energy of the two ions is due to
the collisions of single particle states with the walls of the time-dependent Hartree-Fock potential. This leads to the
randomization of the motion characterized by the distribution of energy among all possible degrees
of freedom of the system. The complete equilibration of the translational kinetic energy among all
possible degrees of freedom is commonly accepted as being the definition of fusion whereas the
incomplete equilibration results in inelastic collisions.

\subsection{\label{sec:geom} TDHF Collision}

\begin{figure}[!htb]
\begin{center}
\includegraphics*[scale=0.45]{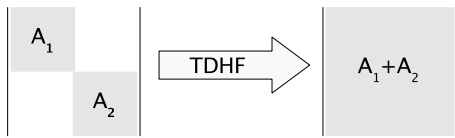}
\caption{\label{fig:tdhf_collision} Schematic illustration of the initial and final many-body states.
The initial state is block diagonal whereas the final state is a full Slater determinant.}
\end{center}
\end{figure}

In TDHF, the initial nuclei are calculated using the static Hartree-Fock (HF) theory. The resulting Slater
determinants for each nucleus comprise the larger Slater determinant describing the colliding
system during the TDHF evolution, as depicted in Fig.~\ref{fig:tdhf_collision}.
Nuclei are assumed
to move on a pure Coulomb trajectory until the initial separation between the nuclear centers used
in TDHF evolution. Using the Coulomb trajectory we compute the relative kinetic energy at this
separation and the associated translational momenta for each nucleus. The nuclei are than boosted
by multiplying the HF states with
\begin{equation}
\Phi _{j}\rightarrow \exp (\imath\mathbf{k}_{j}\cdot \mathbf{R})\Phi _{j}\;,
\end{equation}
where $\Phi _{j}$ is the HF state for nucleus $j$ and $\mathbf{R}$ is the corresponding
center of mass coordinate
\begin{equation}
\mathbf{R}=\frac{1}{A_{j}}\sum _{i=1}^{A_{j}}\mathbf{r}_{i}\;.
\end{equation}
The Galilean invariance of the TDHF equations (discussed below) assures the evolution of
the system without spreading and the conservation of the total energy for the system.
In TDHF, the many-body state remains a Slater determinant at all times. The final state
is a filled determinant, even in the case of two well separated fragments. This phenomenon
is commonly known as the ``cross-channel coupling'' and indicates that it is not possible
to identify the well separated fragments as distinct nuclei since each single particle state
will have components distributed everywhere in the numerical box. In this sense it is
only possible to extract {\it inclusive} (averaged over all states) information from these calculations.

Approximations used in collision geometry include the assumption of an axially symmetric geometry used
in earlier TDHF calculations~\cite{DDKS}. In addition, reflection symmetry with respect to a fixed reaction
plane and z-parity symmetry for identical systems have also been used.
For axially symmetric calculations, non-central collisions were studied using the so called``rotating frame
approximation''~\cite{DK81}.  During the past decade some of these assumptions, specially the axial symmetry
constraint have been relaxed~\cite{KOB97,UO05}. A limited number of comparisons of axially symmetric TDHF calculations
with the corresponding three dimensional calculations are available~\cite{DFFW,FKW,BGK,KOB97}. The three
dimensional calculations show more dissipation as anticipated.

\subsection{\label{sec:skyrme} Effective interaction}

Almost all TDHF calculations have been done using the Skyrme interaction. A variety of calculations have shown
that the TDHF results are very sensitive to the different parametrization of the Skyrme force~\cite{USR,KrLa,RU,US89}.
Fusion behavior is especially sensitive to the effective interaction~\cite{USR}. Some of the assumptions
made in earlier calculations included neglecting the spin-orbit force and assuming spin saturation,
neglect of pairing and the use of the
``filling approximation'' for the occupancy of the last partially filled shell, and the time-reversal invariance
of the single particle Hamiltonian. Most of the earlier TDHF calculations also replaced some of the numerically
difficult terms in the Skyrme interaction with a finite-range Yukawa form~\cite{HN77}, without a new fit to the
nuclear properties.
Previously, we have shown that the inclusion of the spin-orbit interaction
lead to enough additional dissipation to resolve the well known ``fusion window anomaly''(a non-zero lower orbital angular 
momentum limit for fusion)~\cite{USR,KrLa,RU}. Most of the new generation TDHF programs do include at least the
traditional spin-orbit interaction. However, it is well known~\cite{EB75} that the Skyrme energy density functional
also contains terms
which depend on the spin density, $\mathbf{s}$, spin kinetic energy density, $\mathbf{T}$, and the full spin-current pseudotensor,
$\tensor{J}$, as
\begin{equation}
E=\int d^{3}r\;{\cal H}(\rho ,\tau,\mathbf{j},\mathbf{{s}},\mathbf{T},\tensor{J};\mathbf{{r}})\;.
\end{equation}
The time-odd terms ($\mathbf{j}$, $\mathbf{s}$, $\mathbf{T}$) vanish
for static calculations of even-even nuclei, while they are present for odd mass nuclei, in cranking calculations, as well 
as in TDHF. The spin-current pseudotensor, $\tensor{J}$, is time-even and does not vanish for
static calculations of even-even nuclei. However, this terms has not been commonly included in its full extent in the
fitting of the Skyrme parameters due to its numerical complexity (the spin-orbit density $\mathbf{J}$ is the antisymmetric
part of this pseudotensor, and has been included). The inclusions of these terms modifies the Skyrme energy density functional as,
\begin{widetext}
\begin{eqnarray*}
\mathcal{H} = \mathcal{H}_0 &+& \frac{1}{4}t_{0}x_{0}{\mathbf{{s}}}^{2}-\frac{1}{4}t_{0}({\mathbf{{s}}}_{n}^{2}+{\mathbf{{s}}}_{p}^{2})+\frac{1}{24}\rho
^{\alpha }t_{3}x_{3}{\mathbf{{s}}}^{2}-\frac{1}{24}t_{3}\rho ^{\alpha}({\mathbf{{s}}}_{n}^{2}+{\mathbf{{s}}}_{p}^{2}) \\
 &+& \frac{1}{32}(t_{2}+3t_{1})\sum _{q}\mathbf{{s}}_{q}\cdot\nabla^{2}\mathbf{{s}}_{q}-\frac{1}{32}(t_{2}x_{2}-3t_{1}x_{1})\mathbf{{s}}\cdot\nabla ^{2}\mathbf{{s}} \\
 &+& \frac{1}{8}(t_{1}x_{1}+t_{2}x_{2})\left(\mathbf{{s}}\cdot \mathbf{{T}} - \tensor{J}^2\right)
+\frac{1}{8}(t_{2}-t_{1})\sum _{q}\left(\mathbf{{s}}_{q}\cdot\mathbf{{T}}_{q}-\tensor{J}^2_q\right)\\
&-&{\frac{t_{4}}{2}}\sum_{qq'}(1+\delta_{qq'})\left[\mathbf{{s}}_{q}\cdot \nabla\times \mathbf{{j}}_{q'}+\rho_{q}\tensor{\nabla}\cdot\tensor{J}\right]\;,
\end{eqnarray*}
\end{widetext}
where $\mathcal{H}_0$ is the Skyrme energy density functional used in earlier calculations,
with the exception of the spin-orbit
term containing the density $\mathbf{J}$.
The Skyrme energy density functional does remain time-reversal invariant as all the time-odd terms enter
in quadratic form or as linear bi-products. These terms, while required for TDHF to maintain the Galilean invariance of the
collision process~\cite{DD95}, have not been included in TDHF calculations because of numerical difficulty.
Recently, due to renewed efforts towards an improved Skyrme interaction for static nuclear properties, a number
of investigations have focused on identifying the importance and impact of these time-odd terms~\cite{DD95,BHR03,TE05}. It is
clear that they can no longer be neglected in TDHF calculations, at least for preserving Galilean invariance
and assuring that TDHF calculations are truly based on the same static effective interaction, since the most
modern parametrization of the Skyrme force include such terms~\cite{CB98}. Finally, the pairing force has sometimes been
included in TDHF calculations as approximated by BCS type pairing, where BCS equations are solved for the
calculation of the initial static nuclei and the occupation numbers are kept frozen during the time-evolution.
It has previously been argued that, due to the extensive continuum coupling and internal excitations during
the time-evolution, the effects of pairing will quickly wash away~\cite{DK81}, whereas other calculations have
shown stronger pairing correlations~\cite{TU02b}.
There is also the question of handling pairing,
which is inherently related to time-reversal invariance, and the time-reversal breaking terms at the single-particle
level for TDHF calculations.
The study of the importance of pairing interactions during the collision process is still an open question and
can only be properly answered by performing time-dependent Hartree-Fock Bogoliubov (TDHFB) calculations~\cite{TV62}.
Finally, most Skyrme parametrizations include a one-body center-of-mass correction term, which is not included in
generating the initial static solutions for the TDHF evolution.

\subsection{\label{sec:numerical} Numerical approximations}

From the numerical standpoint, new techniques have been developed to handle the solution of the Hartree-Fock 
equations on a space-time lattice. Equations of motion are obtained via the variation of the lattice representations
of the constants of motion, such as the total energy~\cite{DK81,US91a}. In this approach, finite lattice
equations which exactly preserve the constants of motion emerge from the theory in a systematic way.
Most of the earlier numerical calculations have employed low order finite-difference discretization techniques
where the resulting numerical accuracies limited the studies to the gross features of the reaction.
With modern supercomputers it has become feasible to carry out more extensive nuclear
structure and reaction studies employing higher-order discretization techniques, such as fifth and seventh
order finite-difference. Over the last decade we have developed a more modern and advanced technique by discretization
of the energy density functional on a basis-spline collocation lattice, which provides a highly accurate
alternative to the finite-difference method~\cite{US91a,US91b}.


\section{\label{sec:fuscross}fusion cross-sections}

Heavy-ion fusion reactions are a sensitive probe of the size, shape, and structure
of atomic nuclei as well as the collision dynamics. Fusion studies using neutron-rich
nuclei are becoming increasingly available.
In recent experiments with heavy neutron-rich $^{132}Sn$ beams on $^{64}Ni$~\cite{Li03},
enhanced fusion-evaporation cross sections have been observed.
Another experimental frontier is the synthesis of superheavy nuclei in cold and hot fusion
reactions~\cite{Ho02,Og04,Gi03,Mo04,II05}.
Some phenomenological models predict that the fusion cross sections
depend on the heavy-ion interaction potential and on the nuclear form factors in the
vicinity of the Coulomb barrier~\cite{Ba80}.
The more recent theoretical approaches for calculating heavy-ion fusion
cross sections may be grouped into three major categories:
a) barrier penetration models~\cite{Ba80,TB84,RO83,BT98},
b) coupled-channels calculations~\cite{LP84,RP84,HR99,Esb04,Esb05}, and
c) microscopic many-body approaches such as the TDHF method~\cite{Ne82,DDKS,BFH85,USR,KOB97}.

In fusion, the relative kinetic
energy in the entrance channel is entirely converted into internal
excitations of a single well defined compound nucleus. In TDHF theory
the dissipation of the relative kinetic energy into internal excitations is
due to the collisions of the nucleons with the ``walls'' of the
self-consistent mean-field potential. TDHF studies demonstrate that the
randomization of the single-particle motion occurs through repeated
exchange of
nucleons from one nucleus into the other. Consequently, the equilibration of
excitations is very slow and it is sensitive to the details of the
evolution of the shape of the composite system.
This is in contrast to
most classical pictures of nuclear fusion, which generally assume near
instantaneous, isotropic equilibration. Although fusion reactions occur
for light, medium, and heavy-systems there are qualitative and
quantitative differences among these systems. The interpretation of
fusion reactions in terms of a semi-classical TDHF theory exhibits the
best agreement with experiment for the lightest systems,
since here fusion comprises almost the entire reaction cross section.
Since TDHF is a semi-classical theory it is only possible to calculate
fusion cross-sections above the barrier, which is dynamically determined
and may be different than the one calculated using a static two-center model.
Historically, TDHF calculations have been shown to reproduce the general
trends of the observed fusion data \cite{DDKS,BGK,KOB97}.
The TDHF fusion cross-section is calculated using the
sharp-cutoff approximation \cite{BGK}
\begin{equation}
\sigma_f=\frac{\pi\hbar^2}{2\mu E_{cm}}\left(\ell_{max}+1\right)^2\;,
\end{equation}
where $\mu$ is the reduced mass, $E_{cm}$ is the initial center of mass
energy, and the quantity $\ell_{max}$ denotes the maximum
orbital angular momentum for which fusion occurs. Previously, the
above expression for fusion cross-section contained a non-zero lower
limit for orbital angular momentum to accommodate for central transparency
observed for some systems. 
The so called ``fusion-window
anomaly'', which had not been experimentally seen and has been considered to
be the breakdown of the mean-field approach, has been shown to disappear
when the spin-orbit interaction was included in the TDHF calculations~\cite{USR}.
\begin{table}[hbt!] 
\caption{\label{table:threshold_O16}
Calculations of the fusion threshold energy for $^{16}O+^{16}O$ using various
parametrizations of the Skyrme interaction. The subscript $Y$ indicates that the
$\nabla^2\rho$ terms are replaced by a finite-range Yukawa form for computational
reasons. ${\cal T}=0$ indicates no time-reversal symmetry for the interaction was assumed.}
\begin{ruledtabular}
\begin{tabular}{ l c l}
Force & $E_{threshold}$ (MeV)& Comment  \\
\hline
SkII$_Y$    & 68 & Ref. \cite{USR}, 2D, only $\mathbf{j}^2$ \\
SkM$^*_Y$   & 70 & Ref. \cite{USR}, 2D, only $\mathbf{j}^2$ \\
SKM$^*$     & 62 & ${\cal T}=0$, 3D, only $\mathbf{j}^2$\\
SkM$^*$     & 56 & ${\cal T}=0$, 3D, include $\tensor{J}^2$\\
Sly4        & 56 & ${\cal T}=0$, 3D \\
Sly4        & 53 & ${\cal T}=0$, 3D, include $\tensor{J}^2$\\
Sly5        & 55 & ${\cal T}=0$, 3D, fitted with $\tensor{J}^2$
\end{tabular}
\end{ruledtabular}
\end{table}

We have carried out a number of TDHF calculations for the $^{16}O+^{16}O$
system using different parametrizations of the Skyrme force and compared
them to earlier calculations. The calculations were done in an unrestricted
three-dimensional geometry using a  basis-spline collocation lattice of $(-14,+14)^3$ and
a lattice spacing of $1.0$~fm.  The static solutions were obtained using the gradient
iteration method to an energy convergence of $1$ part in $10^{14}$ and
the time evolution used the exponential expansion of the infinitesimal
propagator for up to $15$ terms. Without assuming time reversal invariance
each single particle state is represented by a two-spinor carrying an occupation number of $n=1$.
So, for a single  $^{16}O$ nucleus we have $16$ single-particle states, each having a spin-up
and a spin-down component. The nuclei were initialized assuming that they approach each
other asymptotically on a Coulomb trajectory. The initial separation of the nuclei for TDHF
calculations was taken to be $15$~fm.
Further numerical details and the accuracy of our calculations have been discussed in Ref.~\cite{US91b}.

We first examine the threshold energy for fusion, which is the energy above which
we do not observe fusion but only inelastic collisions. This is done by performing
head-on (zero impact parameter) collisions for various parametrizations of Skyrme the
interaction and compared to earlier TDHF calculations. It should be noted that
for head-on collisions the reduction of TDHF equations to axial-symmetry is almost
exact.  The results are tabulated in Table~\ref{table:threshold_O16} along with
comments indicating the details of the force selection in each case. Since the
inclusion of the spin-orbit interaction was found to have a profound impact on
these results~\cite{USR} we do not discuss prior results that do not include this
contribution. The first two threshold values denote the calculations done using
axially symmetric geometry and the Yukawa finite-range approximation for the Skyrme
parametrizations SkII~\cite{VB72} and SkM$^{*}$~\cite{BQ82}.
Traditionally, all TDHF calculations included the time-odd current $\mathbf{j}$
appearing in combination $(\rho\tau-\mathbf{j}^2)$.
\begin{table}[hbt!] 
\caption{\label{table:fusion_O16}
Calculations of the fusion cross section for $^{16}O+^{16}O$ using various
parametrizations of the Skyrme interaction. }
\begin{ruledtabular}
\begin{tabular}{ l c}
Force                     & $\sigma_{fusion}$ (mb) \\
\hline
SkII$_Y$ \cite{USR}   & 1694 \\
SkM$^*_Y$\cite{USR}   & 1822 \\
SkM$^*$                   & 1368 \\
Sly5                      & 1347 \\
Experiment                & 1075
\end{tabular}
\end{ruledtabular}
\end{table}
The next value is the same
calculation performed using the exact form of the SkM$^{*}$ interaction and in
three-dimension. As we see, the threshold energy is reduced by $8$~MeV. Since,
axially symmetric geometry is almost exact for head-on collisions this difference
could be largely attributed to the incorrect surface behavior generated by the
Yukawa approximation and perhaps to substantially improved numerical accuracy.
The next four threshold values include all of the time-odd terms in the Skyrme interaction.
However, there is still an unresolved issue regarding the terms containing the time-even pseudotensor
$\tensor{J}$. This term is non-zero for static calculations but
has not been fully included in most fits for the Skyrme interaction. On the other hand
it may be necessary to maintain the Galilean invariance of the TDHF evolution.
Repeating the same calculation for SkM$^{*}$ but including all of the terms in
the Skyrme interaction results in a reduction of the threshold energy by another
$6$~MeV. Finally, we have performed calculations with more modern Skyrme forces,
SLy4 and SLy5 \cite{CB98}. The parametrization SLy4 does not include the
$\tensor{J}^2$ contribution to the Skyrme energy density
functional. The inclusion of this term results in a $3$~MeV reduction in threshold
energy. This is interesting because the contribution of this term to the binding
energy of the $^{16}O$ nucleus is on the order of a few tens of keV. The last row
of Table~\ref{table:threshold_O16} shows the result for the SLy5 parametrization,
which includes the $\tensor{J}^2$ contribution in determining
the force parameters. We can conclude that, despite small differences, most modern forces
seem to yield threshold energies that are in agreement with each other.
\begin{figure}[!htb]
\begin{center}
\includegraphics*[scale=0.35]{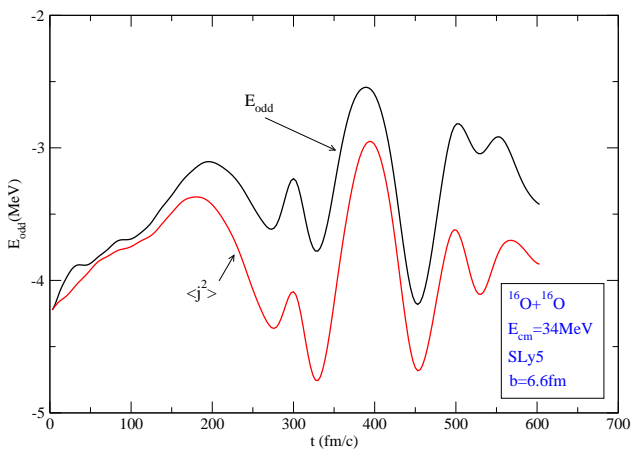}
\caption{\label{fig:time-odd} (Color online) The contribution of time-odd terms to the total energy is
plotted as a function of time for the $^{16}O+^{16}O$ system at $b=6.6$~fm, using the
SLy5 interaction. On the same figure we also separately show the contribution arising from the time-odd
$\mathbf{j}^2$ term, which was also present in earlier TDHF calculations.}
\end{center}
\end{figure}

We have also performed fusion calculations for the $^{16}O+^{16}O$ system at a
center of mass energy of $34$~MeV. The reasons for choosing this particular collision
energy is due to the availability of older calculations and data, as well as
increased sensitivity to the details of the nuclear interaction~\cite{USR} for
this relatively high energy collision.
The results are
tabulated in Table~\ref{table:fusion_O16} for various parametrizations of the Skyrme
force. The maximum impact parameter for fusion was searched in $1.0$~fm intervals
until no fusion was observed, which was then followed by a more precise search in
intervals of $0.05-0.1$~fm. Maximum impact parameters were found to be $6.65$~fm
and $6.60$~fm for SkM$^{*}$ and SLy5, respectively.
Again, we see substantial improvement with the more
modern Skyrme forces when no approximation in geometry and interaction is used.
The reduction in the total fusion cross section of $500$~mb is in the right direction
but still overestimates the experimental cross-section~\cite{BT79} by about $25$\%.
In order to better understand the contribution of the various new terms contained
in the time-odd part of the interaction we have plotted the total energy arising
from the time-odd part of the Skyrme energy density functional in Fig.\ref{fig:time-odd}.
On the same figure we also separately show the contribution arising from the time-odd
$\mathbf{j}^2$ term, which was present in earlier TDHF calculations. As we see,
the total contribution  traces the behavior of the contribution from the $\mathbf{j}^2$ term
with a slight overall shift.
\begin{figure}[!htb]
\begin{center}
\includegraphics*[scale=0.35]{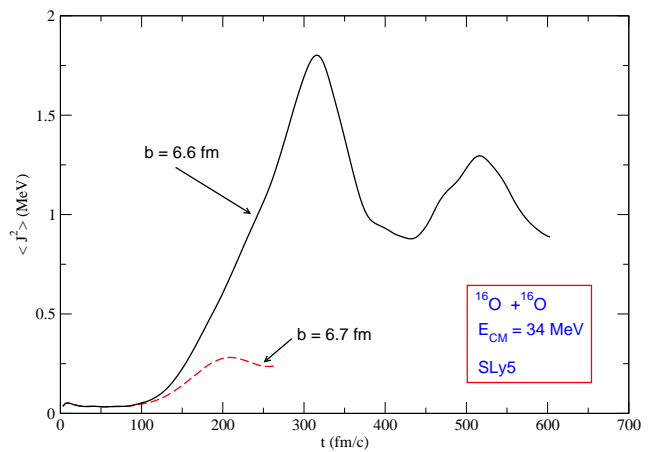}
\caption{\label{fig:J2} (Color online) The contribution of the time-even pseudotensor term 
to the total energy is plotted as a function of time for the $^{16}O+^{16}O$ system at $b=6.6$~fm
and $b=6.7$~fm,
using the SLy5 interaction.}
\end{center}
\end{figure}
\begin{figure*}[!htb]
\begin{center}
\includegraphics*[scale=0.32]{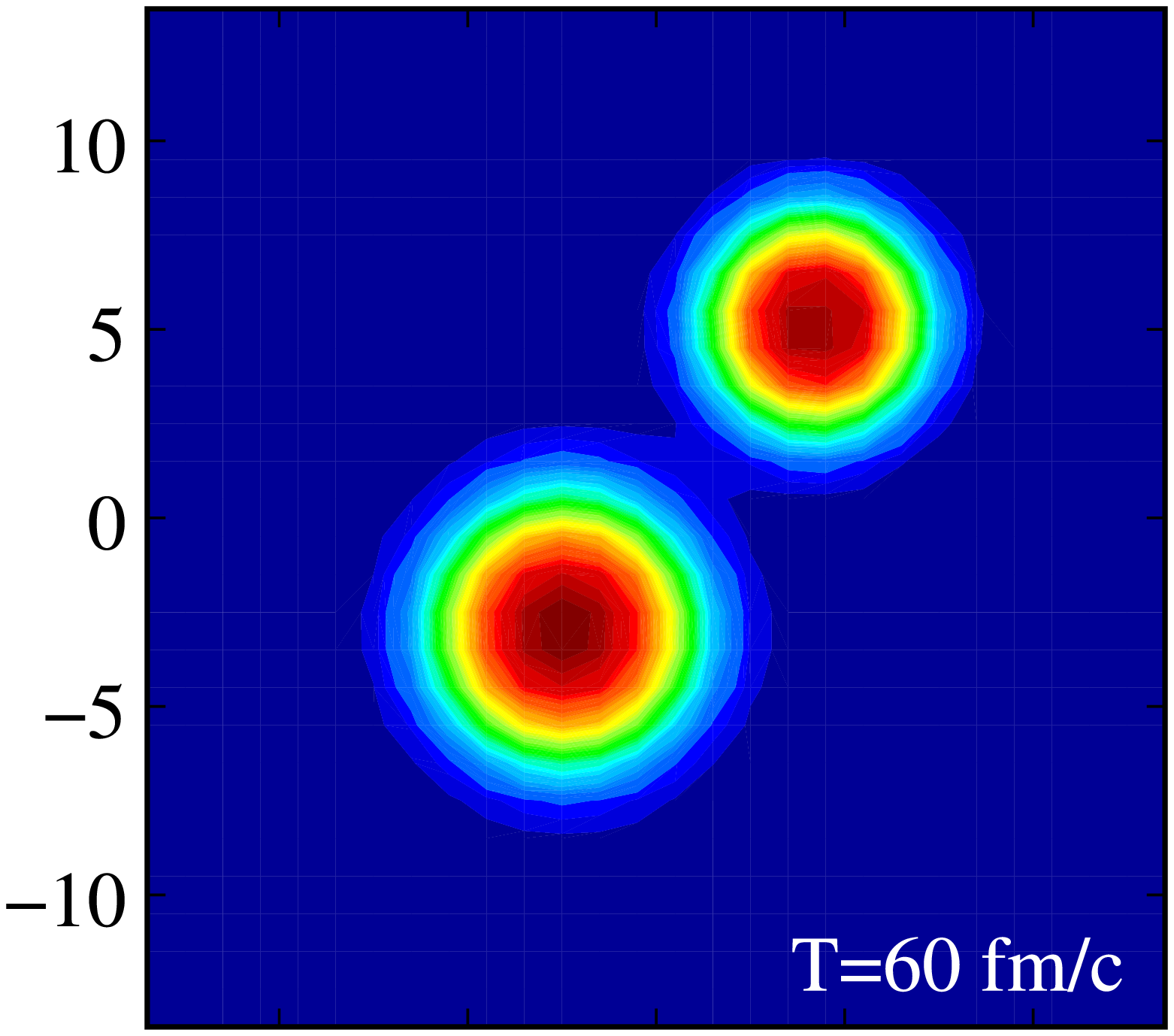}\hspace{-0.05in}
\includegraphics*[scale=0.32]{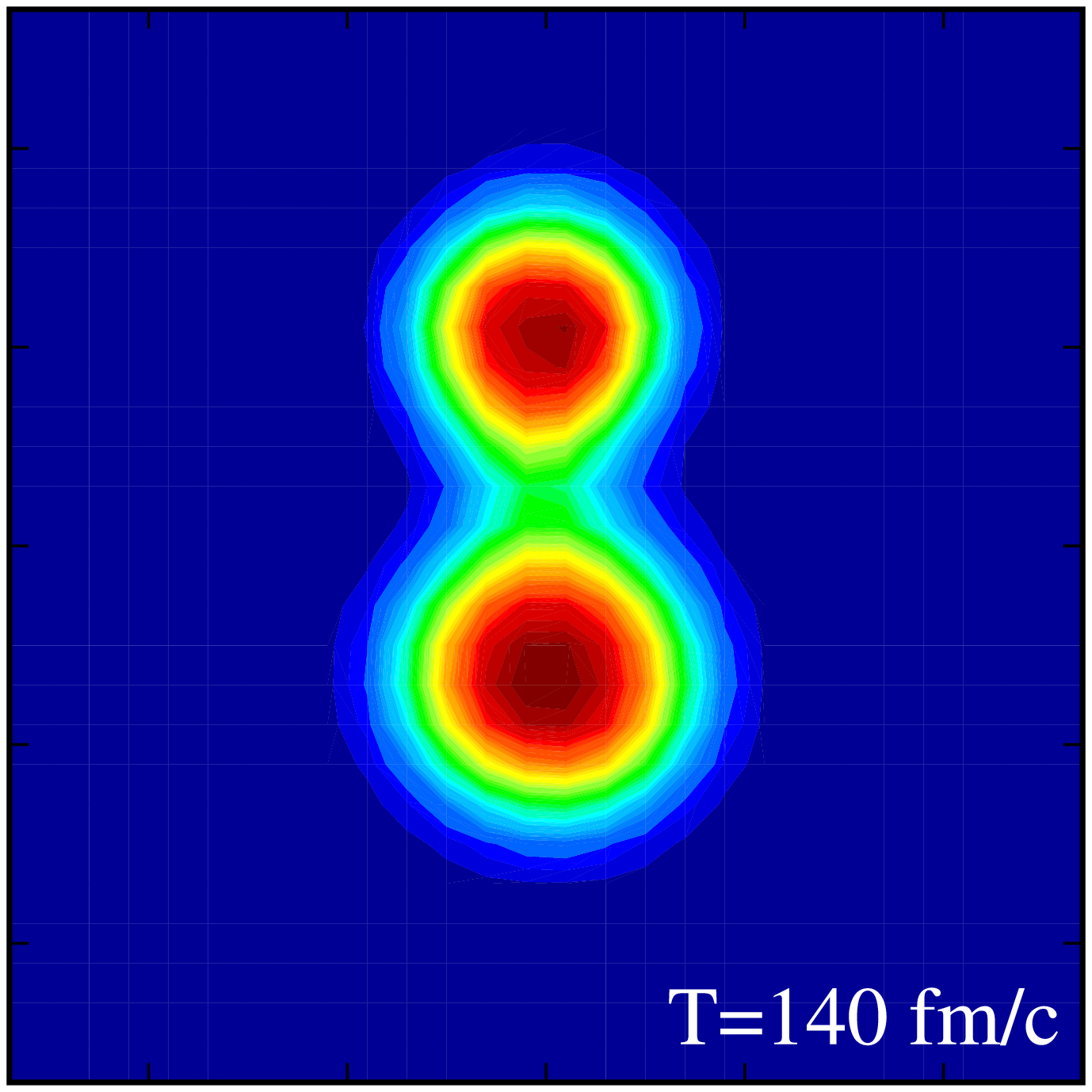}\hspace{-0.05in}
\includegraphics*[scale=0.32]{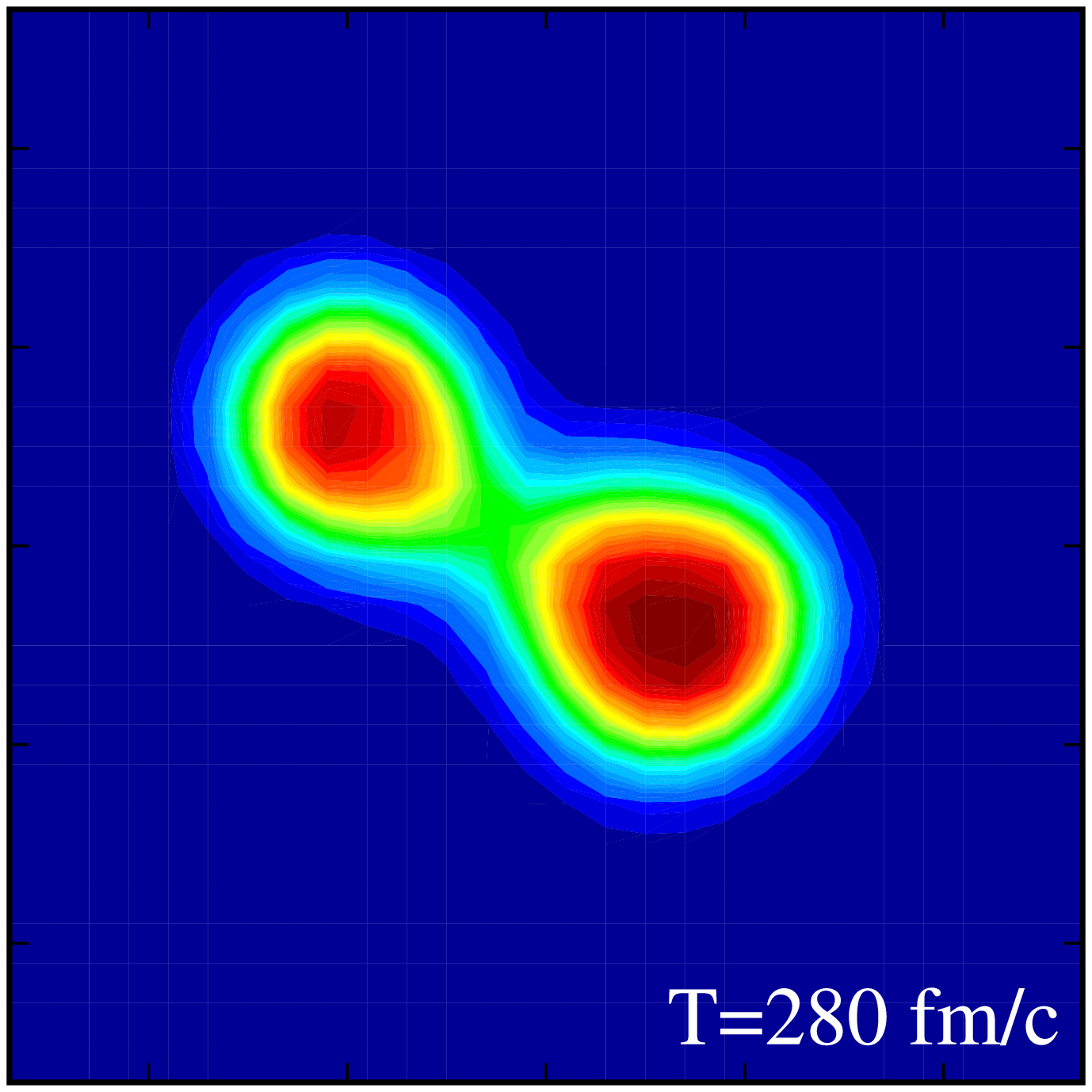}\\ \vspace{-0.05in} \hspace{0.58in}
\includegraphics*[scale=0.32]{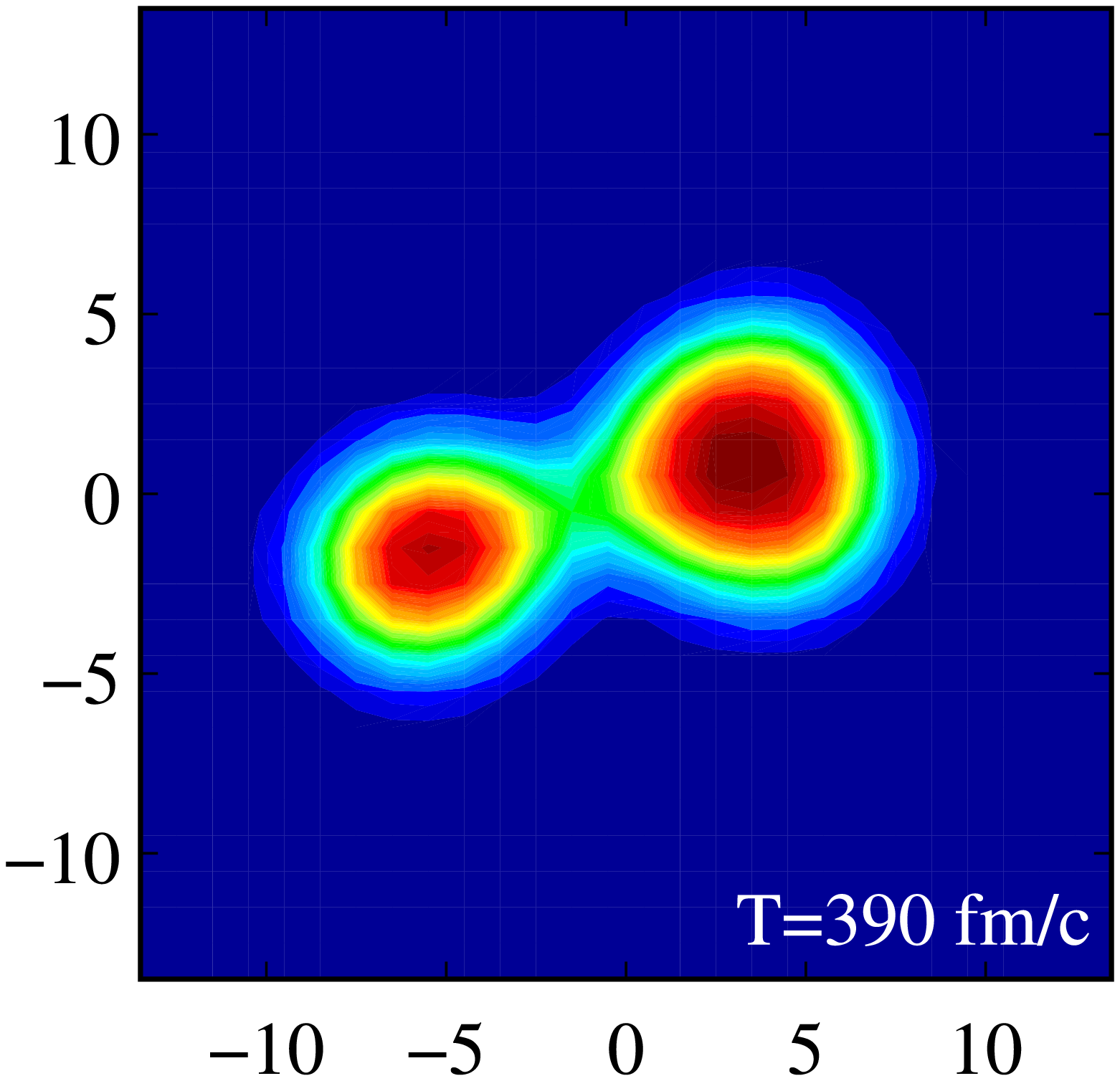}\hspace{-0.05in}
\includegraphics*[scale=0.32]{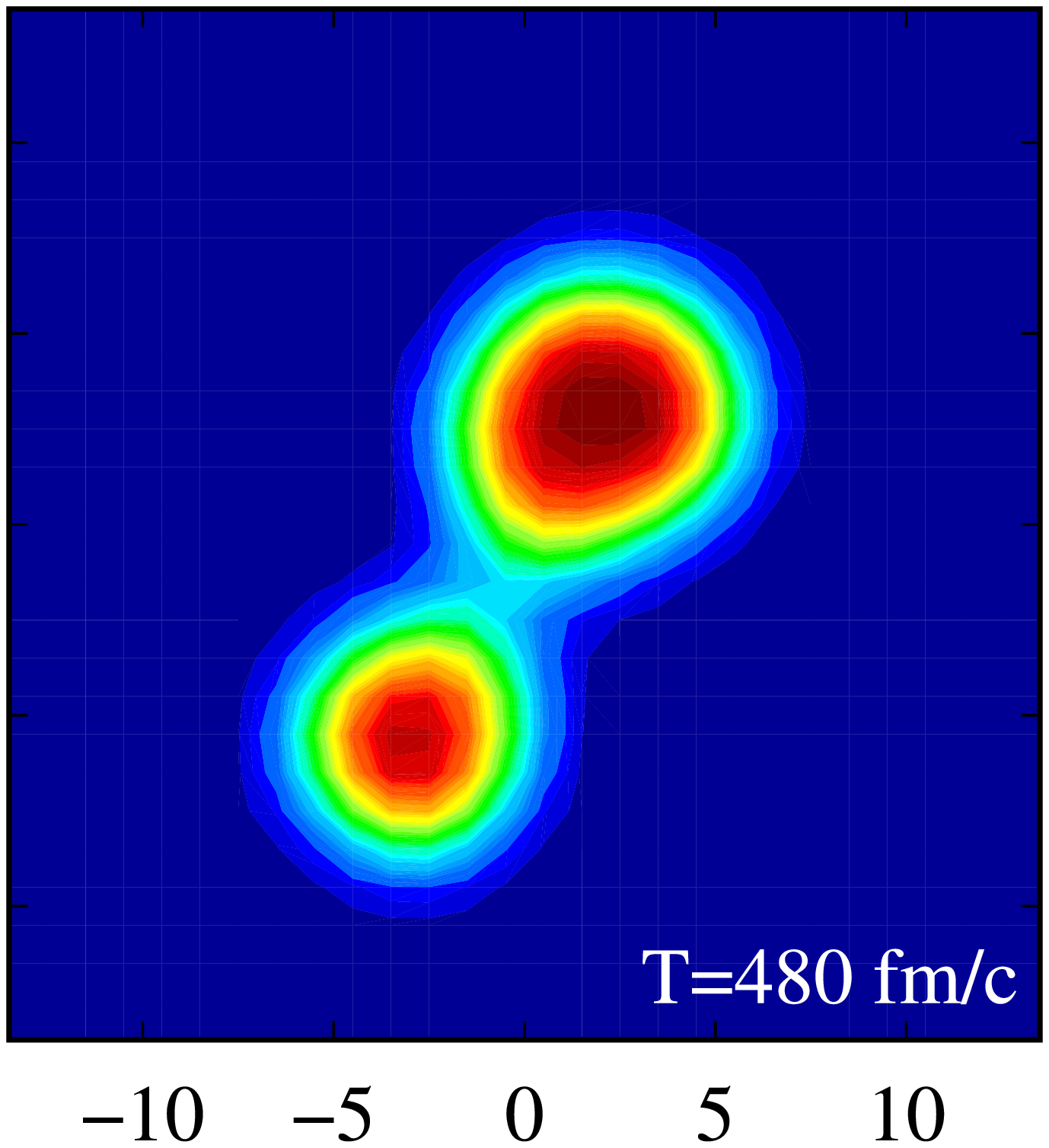}\hspace{-0.05in}
\includegraphics*[scale=0.32]{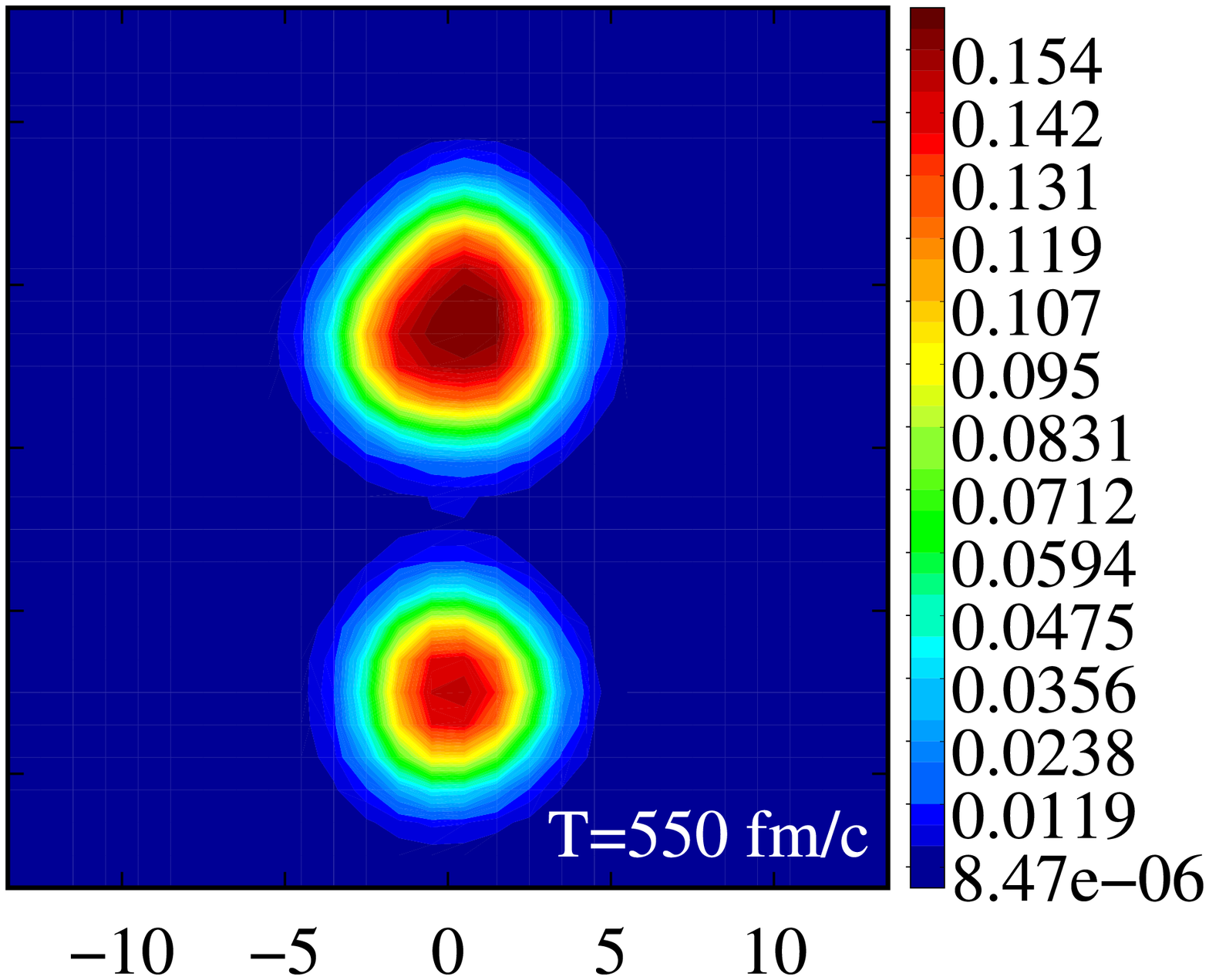}
\caption{\label{fig:time-evol} (Color online) TDHF time-evolution
for the $^{16}O+^{28}O$ collision at an impact parameter of $b=7.6$~fm, just above the fusion
region, using the SLy5 interaction. The initial energy is $E_{cm}=43$~MeV.}
\end{center}
\end{figure*}
In Fig.~\ref{fig:J2} we plot the time-evolution of the contribution to the total energy arising from the 
$\tensor{J}^2$ term. We have plotted two impact parameters, first for
$b=6.6$~fm, where fusion occurs, and the second at $b=6.7$~fm, where there is no fusion.
We observe that when the nuclei are far apart the contribution is small and it grows as nuclei
approach each other. However, we see a major difference between the case for which there is
fusion and the case were no fusion occurs. In the former, the contribution rises rapidly and
reaches its maximum around the time of greatest overlap during the neck formation. It then
remains significant during the formation of the compound system. In the latter case the
contribution remains small throughout the collision process, and finally the  two fragments, albeit excited, 
come apart and move away from each other. This is an interesting result since this term does
not seem to make a major contribution to the binding energy and does not significantly
alter the parameters of the Skyrme force when it is included in the fits~\cite{CB98}.
We can conclude from this that the nuclear properties used in fitting the parameters of
the Skyrme force is not triggering the physical significance of this term.

We have repeated these calculations for the collision of $^{16}O$ with the neutron-rich
$^{28}O$ nucleus at $E_{cm}=43$~MeV, which maintains the same initial velocity as in the  
$^{16}O+^{16}O$ case. The maximum impact parameter for fusion is found to be $7.5$~fm,
which results in a cross-section of approximately $1767$~mb. This value scales well with
the mass number and does not indicate an enhanced fusion cross-section for this neutron-rich
system. Of course, the energy is relatively high and such enhancements may be seen at lower
energies or below the barrier. In Fig.~\ref{fig:time-evol} we show this collision for
$b=7.6$~fm, for which there is no fusion. In this deep-inelastic collision the final 
translational energy of the separating ions is about $20$~MeV, indicating that $23$~MeV
was utilized for internal excitations. The final fragments, besides being highly excited,
show an exchange of approximately two protons and a neutron to the heavy fragment.
The analysis of the various contributions arising from the terms in the Skyrme interaction
show a similar behavior to the $^{16}O+^{16}O$ system. 

\section{Summary and Outlook}

The evolution of the TDHF studies of heavy ion reactions is marked
by increasingly sophisticated calculations, trying to eliminate as many
of the assumed symmetries as possible. This progress has closely paralleled the
advances in computer technology.

We have presented calculations using a new generation TDHF program, 
which makes no assumptions regarding the collision geometry nor the
Skyrme interaction, and uses advanced numerical methods for improved
accuracy. We have compared the new results with earlier TDHF calculations
and analyzed the influence of the new terms in the effective interaction,
specifically the new time-odd terms and the spin-current pseudotensor contribution.
In general, unrestricted calculations and new Skyrme parametrizations lead to
substantial improvements of fusion results. The substantially different results obtained
by earlier parametrizations of the Skyrme force seems to have converged to
very similar outcomes for the modern parametrizations, a sign of major strides
made in improving the Skyrme interaction. On the other hand,
we find that some of  the new terms make an appreciable contribution during the
dynamical evolution, while being absent or minimally important for the static
calculations. This suggest that improvements to the Skyrme parametrization
are still possible by incorporating dynamical features into the fitting process,
along the lines of Refs.\cite{DD95,RF95}.
It seems as if the mean-field approach has not yet been fully exhausted, and improved
TDHF calculations may display
more realistic features for heavy-ion collisions at low and medium energies.


\begin{acknowledgments}
This work has been supported by the U.S. Department of Energy under grant No.
DE-FG02-96ER40963 with Vanderbilt University. Some of the numerical calculations
were carried out at the IBM-RS/6000 SP supercomputer of the National Energy Research
Scientific Computing Center which is supported by the Office of Science of the
U.S. Department of Energy.
\end{acknowledgments}




\end{document}